\documentclass[12pt,onecolumn,tightenlines,
superscriptaddress,nofootinbib,preprintnumbers]{revtex4-2}
\usepackage[english]{babel}
\usepackage{amsthm}
\usepackage{comment}
\usepackage{dsfont}
\usepackage{bbold}

\usepackage[utf8]{inputenc}

\usepackage{csquotes}

\usepackage{xspace}
\usepackage{graphicx}
\usepackage{cancel} 
\usepackage{multirow}
\graphicspath{{./figures/}}
\usepackage[caption=false]{subfig}

\usepackage{tikz}
\usetikzlibrary{
  arrows.meta,
  calc,
  decorations.markings,
  fit,
  matrix,
  positioning,
  shapes,
  matrix,decorations.pathreplacing
}

\usepackage[letterpaper,top=2cm,bottom=2cm,left=3cm,right=3cm,marginparwidth=1.75cm]{geometry}

\usepackage{amsmath}
\numberwithin{equation}{section}
\usepackage{mathtools}
\usepackage{bm} % bold math symbols
\usepackage[colorlinks=true]{hyperref}
\hypersetup{
  linkcolor=blue,
  citecolor=blue,
  filecolor=red,
  urlcolor=blue,
  menucolor=red,
  runcolor=red
}

\def\eps{\epsilon}
\def\eqn#1{eq.~\eqref{#1}}
\def\eqns#1#2{eqs.~\eqref{#1} and~\eqref{#2}}
\def\sect#1{Sect.~\ref{#1}}
\def\spa#1.#2{\left\langle#1\,#2\right\rangle}
\def\spb#1.#2{\left[#1\,#2\right]}

\begin{document}
\preprint{USTC-ICTS/PCFT-25-30, MPP-2025-151}
\title{Singularity-Free Feynman Integral Bases}

\author{Stefano De Angelis}
\affiliation{Institut de Physique Théorique, CEA, CNRS, Université Paris--Saclay, F--91191 Gif-sur-Yvette cedex, France\\
\textsf{\rm\sf Stefano.De-Angelis@ipht.fr}}

\author{David A. Kosower}
\affiliation{Institut de Physique Théorique, CEA, CNRS, Université Paris--Saclay, F--91191 Gif-sur-Yvette cedex, France\\
\textsf{\rm\sf David.Kosower@ipht.fr}}

\author{Rourou Ma}
\affiliation{Interdisciplinary Center for Theoretical Study, University of Science and Technology of China, Hefei, Anhui 230026, China}
\affiliation{Max-Planck-Institut f\"ur Physik,  Werner-Heisenberg-Institut, Boltzmannstraße 8, 85748 Garching, Germany\\
\textsf{\rm\sf Marr21@mail.ustc.edu.cn}}

\author{Zihao Wu}
\affiliation{Hangzhou Institute for Advanced Study, University of Chinese Academy of Sciences (HIAS, UCAS)\\
\textsf{\rm\sf wuzihao@mail.ustc.edu.cn}}

\author{Yang Zhang}
\affiliation{Interdisciplinary Center for Theoretical Study, University of Science and Technology of China, Hefei, Anhui 230026, China}
\affiliation{Peng Huanwu Center for Fundamental Theory, Hefei, Anhui 230026, China}
\affiliation{Center for High Energy Physics, Peking University, Beijing 100871, People’s Republic
of China\\
\textsf{\rm\sf yzhphy@ustc.edu.cn}}

%\author[b,c]{Rourou Ma}
%\author[d]{Zihao Wu}
%\author[b,e]{Yang Zhang}

%\emailAdd{Marr21@mail.ustc.edu.cn}
%\emailAdd{wuzihao@mail.ustc.edu.cn}
%\emailAdd{yzhphy@ustc.edu.cn}

\date{\today}

\begin{abstract}
Standard integration-by-parts (IBP) reduction methods
typically yield Feynman integral bases where the reduction
of some integrals gives rise to coefficients singular as the dimensional regulator $\eps\rightarrow 0$.
These singular coefficients can also appear in scattering amplitudes, obscuring their structure, and rendering their evaluation more complicated.
We investigate the use of bases in which the reduction
of any integral is free of singular coefficients. 
We present two general algorithms for constructing 
such bases.  The first is based on sequential $D=4$ IBP
reduction.  It constructs a basis iteratively by projecting
onto the finite part of the set of IBP relations.
The second algorithm performs Gaussian elimination within
a local ring forbidding division by $\eps$ while 
permitting division by polynomials in $\eps$ finite
at $\eps=0$.
We study the application of both algorithms to a pair of
two-loop examples, the planar and nonplanar double-box
families of integrals.  We also explore the incorporation
of finite Feynman integrals into these bases.  In one
example, the resulting basis provides a simpler and more
compact representation of a scattering amplitude.
\end{abstract}

\maketitle
\newpage

\section{Introduction}

Scattering amplitudes beyond tree level can be
expressed as sums over Feynman integrals with coefficients
rational in kinematic quantities.   
The complete set of Feynman integrals relevant to any
given scattering amplitude is subject to a large set of
linear identities. We can obtain these identities
systematically using integration-by-parts (IBP)
techniques~\cite{Tkachov:1981wb,Chetyrkin:1981qh}.
Eliminating the identities reduces the
set of all Feynman integrals to a basis; but a given
basis is of course not unique.  Some bases are better
than others.  Indeed, different bases may be adapted
to different purposes: generating and solving
a set of differential equations to obtain explicit
expressions for the integrals; finding a compact 
representation for the amplitude (and thereby minimizing
computational effort in evaluating it); or manifesting
certain properties in special limits.  

In practical applications, the systems of equations
that emerge from IBPs are large, and require specialized
codes to solve them.  
Solution methods for these systems of equations
have been a subject of ongoing research since the pioneering
algorithm of Laporta~\cite{Laporta:2000dsw}.  
Researchers have developed a variety of general-purpose
software packages implementing the algorithm and related
ideas over the years, 
including \textsf{AIR}~\cite{Anastasiou:2004vj}, 
\textsf{FIRE}~\cite{Smirnov:2008iw,Smirnov:2013dia,%
Smirnov:2014hma,Smirnov:2019qkx}, 
\textsf{LiteRed}~\cite{Lee:2012cn,Lee:2013mka}, 
\textsf{Reduze}~\cite{Studerus:2009ye,vonManteuffel:2012np}, 
\textsf{Kira}~\cite{Maierhofer:2017gsa,Maierhofer:2018gpa,%
Maierhofer:2019goc,Klappert:2020nbg,Lange:2025fba}, 
\textsf{Blade}~\cite{Guan:2019bcx,Liu:2021wks,Guan:2024byi}, 
and \textsf{NeatIBP}~\cite{Wu:2023upw,Wu:2025aeg}.
In related developments, new finite-field
techniques for functional
reconstruction and solving large systems of linear equations
based on computation in finite fields have helped increase
the power of IBP solvers~\cite{vonManteuffel:2014ixa}.
These techniques have been implemented in specialized 
tools such as \textsf{FiniteFlow}~\cite{Peraro:2016wsq}, 
\textsf{FireFly}~\cite{Klappert:2019emp,Klappert:2020aqs}, 
and \textsf{RaTracer}~\cite{Magerya:2022hvj}.
Other recent work has examined choosing master integrals to
avoid mixed poles involving both the dimensional regulator
and the Mandelstam 
invariants~\cite{Usovitsch:2020jrk,Smirnov:2020quc}, 
and ways of simplifying IBP coefficients~\cite{Boehm:2020ijp,Bendle:2021ueg}.
Researchers have also developed algorithms to find 
uniform-transcendentality 
bases~\cite{Kotikov:2012ac,Lee:2014ioa,Argeri:2014qva,%
Dlapa:2020cwj,Dlapa:2022wdu}, useful in setting up systems of
differential equations for integrals.  An approach quite
distinct from IBP systems --- intersection theory --- has 
also provided new insights into reduction of Feynman 
integrals~\cite{Mastrolia:2018uzb,Chestnov:2022alh,%
Brunello:2024tqf}.

With a basis in hand for a specific process at
a desired loop order, any Feynman
integral that may arise in computing
the amplitude can be expressed in terms of it.  
Canonical 
bases~\cite{Henn:2013pwa,Henn:2014qga},
for example, are 
well-suited to computing the integrals via 
the method of differential equations 
\cite{Kotikov:1990kg,Bern:1993kr,Remiddi:1997ny,%
Gehrmann:1999as}.  Such a basis, however,
may not necessarily be best-suited
to obtaining a compact representation for scattering amplitudes.
Unpleasant features in standard bases 
may include the presence of
integral coefficients singular in the dimensional regulator
$\eps$, or cancellations of divergences between different
integrals. These may arise in the reduction of some integrals,
and may also be present in the expression for scattering 
amplitudes.

In this article, we study the utility of
\textit{singularity-free} bases.  These are bases
where no explicit singularities in $\eps$ appear in 
the coefficients
of basis integrals in the reduction
of any integral, and no explicit singularities in $\eps$
are present in the definition of any basis integral.  
All singularities in $\eps$ then arise solely from the
loop integrations.
Such a basis is expected to improve numerical evaluation 
and stability~\cite{Chetyrkin:2006dh}.  We also conjecture
that it can simplify expressions for physical quantities.

We present two algorithms for
obtaining such bases.  We also examine the possibility
of combining the criterion of singularity freedom with
the use of locally finite Feynman integrals.  Such
integrals were considered in 
refs.~\cite{vonManteuffel:2014qoa,vonManteuffel:2015gxa,%
Agarwal:2020dye},
and most recently by Gambuti, Novichkov, Tancredi, and one 
of the authors (GKNT)~\cite{Gambuti:2023eqh}.
These authors presented a comprehensive approach to finding
and organizing the set of locally finite Feynman integrals.
The GKNT approach relies on classifying singular surfaces in
loop-momentum space using the Landau equations.  De~La~Cruz,
Novichkov, and one of the authors~\cite{delaCruz:2024xsm} presented an equivalent
approach relying on Newton polytopes.  The resulting sets of
finite integrals are not independent under IBPs, but a
subset could be included as basis integrals.  Including finite integrals in the basis may simplify expressions for scattering amplitudes and reduce the extent of cancellations between different divergent terms.

In the next section, we review the appearance of
singular coefficients in integral reduction.
In \sect{SingularityFreeAlgorithms}, we give two
algorithms for obtaining a singularity-free basis
starting with a set of integrals and IBP relations
between them.  In \sect{ExamplesSection}, 
we explore two examples,
that of the planar and nonplanar double box integrals.
Our conclusions are in \sect{ConclusionsSection}.  Two
appendices give technical details on one of the algorithms,
and the coefficients of the subleading-color terms in
the all-plus two-loop amplitude in a modified basis.

\section{Singular Coefficients in Integral Reduction}
\label{SingularCoefficientsSection}

\def\loopm{\ell}
\def\Den{\mathcal{D}}
In general, we will want to consider Feynman integrals 
with nontrivial numerators,
\begin{equation}
  \label{eq:feynman_integral}
I[\mathcal{N}(\ell_i)] =
  \int \prod_{i=1}^{L} \mathrm{d}^D \ell_i \,
  \frac{\mathcal{N}(\ell_i)}{\Den_1 \cdots \Den_E}\,.
\end{equation}
\def\db{I_{\textrm{DB}}}
The simplest nontrivial examples of singular coefficients
arise at two loops, and we will focus on two-loop integrals
for the explicit examples in this article.
Let's look at the planar double box as a first example, with
\begin{equation}
\begin{aligned}
\Den_1&=\loopm_1^2,\quad
\Den_2=(\loopm_1-k_1)^2,\quad
\Den_3=(\loopm_1-K_{12})^2,\quad
\Den_4=(\loopm_1-\loopm_2)^2,
 \\ 
\Den_5&=(\loopm_2-K_{12})^2,\quad 
\Den_6=\loopm_2^2,\quad 
\Den_7=(\loopm_2-K_{123})^2\,,
\end{aligned}
\label{DoubleBoxDenominators}
\end{equation}
where $K_{i\cdots j}=k_i+\cdots+k_j$.  We denote
integrals with these propagators by the subscript `DB'.
\def\Gram#1#2{G\begin{pmatrix}#1\\#2\\ \end{pmatrix}}
\def\GramOne#1{G\begin{pmatrix}#1\end{pmatrix}}
Define the Gram determinant,
\begin{equation}
	\Gram{q_1 &,\dots, & q_m}{p_1 &,\dots, & p_m}
    = \det (2 q_i \cdot p_j)\, ,
\end{equation} 
where we list only one series of arguments if the two
sets are identical.  We represent momenta in the arguments
by their indices alone, so that for example,
\begin{equation}
    \Gram{\loopm_1&2&3}{1&2&3} \equiv 
    \Gram{\loopm_1&k_2&k_3}{k_1&k_2&k_3}\,.
\end{equation}

\def\Ord{\mathcal{O}}
\def\BasisI#1{B_{#1}}
\def\ISP{\textrm{ISP}}
Consider two different but standard choices of basis,
the Laporta basis \cite{Laporta:2000dsw} and 
a uniform-transcendentality one (useful for
differential equations~\cite{Henn:2013pwa} after
adjusting overall normalizations to obtain a canonical basis).
We also have two irreducible scalar products
(ISPs),
\begin{equation}
\label{PlanarDoubleBoxISPs}
    \ISP_1 = (\loopm_1+k_4)^2\,,\qquad
    \ISP_2 = (\loopm_2+k_1)^2\,.
\end{equation}
The Laporta basis for $s$-channel integrals has the
following elements,
\begin{equation}
\begin{aligned}
\{
&\BasisI{0,0,1,0,0,1,1,0,0},\BasisI{0,1,0,0,1,0,1,0,0},
\BasisI{0,1,0,1,0,1,1,0,0},\BasisI{0,1,0,1,1,1,1,0,0},\\
&\BasisI{0,1,1,0,1,1,1,0,0},\BasisI{1,0,1,1,0,1,0,0,0},
\BasisI{1,1,1,1,1,1,1,-1,0},\BasisI{1,1,1,1,1,1,1,0,0}
\}\,,
\end{aligned}
\label{LaportaBasis}
\end{equation}
where the indices correspond as usual to the exponents
of the denominators in \eqn{DoubleBoxDenominators},
followed by the negative of the exponents of the
two ISPs in the numerator~\eqref{PlanarDoubleBoxISPs}. 
A possible uniform-transcendentality basis for 
$s$-channel integrals
has the following elements \cite{Henn:2013pwa},
\begin{equation}
\begin{aligned}
\{
&\BasisI{0,0,2,0,0,1,2,0,0},\BasisI{0,1,0,1,0,1,2,0,0},
\BasisI{0,1,1,0,1,1,1,0,0},\BasisI{0,2,0,0,1,0,2,0,0},\\
&\BasisI{2,0,1,1,0,2,0,0,0},\BasisI{1,1,1,0,1,0,2,0,0},
\BasisI{1,1,1,1,1,1,1,0,0},\BasisI{1,1,1,1,1,1,1,0,-1}
\}\,.
\end{aligned}
\label{UTBasis}
\end{equation}

In the Laporta basis~\eqref{LaportaBasis}, 
we have the following representations for two example integrals,
\begin{equation}
\begin{aligned}
\db[\Den_6^2] &= 
%%%%% begin : Idb1
\frac{3  (1 - 3 \eps)(2 - 3 \eps)}{2 s t \eps^3} 
    \BasisI{0,0,1,0,0,1,1,0,0}
%\\&
+\frac{(3 s+t \eps )}{t\, \eps }\BasisI{0,1,0,1,1,1,1,0,0}
\\&
+\frac{(3 (s+t)-2 t\eps)}{t (1-2 \eps)}
   \BasisI{0,1,1,0,1,1,1,0,0}
-\frac{3 (1-3 \eps)}{2 t\,\eps ^2} \BasisI{0,1,0,1,0,1,1,0,0}
\\&
+\frac{ (1-3 \eps)(2-3 \eps)(3-2 \eps) }{2 t^2\, \eps ^3}
\BasisI{0,1,0,0,1,0,1,0,0}
%%%%% end : Idb1
\,,\\
\db\biggl[\Gram{\loopm_1&1&2}{\loopm_2&3&4}\biggr] &= 
%%%%% begin : Idb2
\frac{s^2}{8}  \BasisI{1,1,1,1,1,1,1,-1,0}
-\frac{(1-\eps) (1-3 \eps) (2-3 \eps) }
{2 s \eps^3}\BasisI{0,0,1,0,0,1,1,0,0}
\\&
+\frac{s+t}{2}\BasisI{0,1,1,0,1,1,1,0,0}
+\frac{(1-2 \eps) (1-3 \eps)(2-3 \eps) }{4 t\, \eps ^3}
\BasisI{0,1,0,0,1,0,1,0,0}\hspace{-4mm}
\\&-\frac{5 (1-2 \epsilon) (1-3 \epsilon)}
{16 \eps^2}\BasisI{0,1,0,1,0,1,1,0,0}
-\frac{(1-\epsilon) (1-2 \epsilon)}{4 \eps ^2}
\BasisI{1,0,1,1,0,1,0,0,0}
%%%%% end : Idb1
\,.\hspace{-10mm}
\end{aligned}    
\label{LaportaBasisExamples}
\end{equation}
The first integral diverges as $1/\eps^4$ as 
$\eps\rightarrow0$, but its
reduction has singular coefficients.  
The second integral is in fact finite,
but this emerges only through cancellations and
in the presence of divergent coefficients.

The expressions in the uniform-transcendentality
basis~\eqref{UTBasis} are simpler,
\begin{equation}
\begin{aligned}
\db[\Den_6^2] &= 
%%%%% begin : Idb3
-\frac{3 s }{4 t \eps ^2(1-2\eps)}
\BasisI{0,0,2,0,0,1,2,0,0}
+\frac{ (3 (s+t)-2 t\eps)}{t\, (1-2 \eps)}
\BasisI{0,1,1,0,1,1,1,0,0}
\hspace*{-15mm}\\&
-\frac{ (3 s+t \eps )}{3 \eps (1-2 \eps)}
\BasisI{1,1,1,0,1,0,2,0,0}
-\frac{(3-2 \eps ) }{4 \eps ^2 (1-2 \eps)}
\BasisI{0,2,0,0,1,0,2,0,0}
\\&-\frac{\BasisI{0,1,0,1,0,1,2,0,0}}{1-2 \eps}
%%%%% end : Idb3
\,,
\\
\db\biggl[\Gram{\loopm_1&1&2}{\loopm_2&3&4}\biggr] &= 
%%%%% begin : Idb4
-\frac{s^2\, (1-\eps) }{4 \eps ^2\, (1-2 \eps)}
\BasisI{2,0,1,1,0,2,0,0,0}
+\frac{s^2}{8}  \BasisI{1,1,1,1,1,1,1,0,-1}
\\&
+\frac{s+t}{2}\BasisI{0,1,1,0,1,1,1,0,0}
+\frac{s (1-\eps)}{4 \eps ^2\, (1-2 \eps)}
\BasisI{0,0,2,0,0,1,2,0,0}
\\&+\frac{5 s}{8 \eps }\BasisI{0,1,0,1,0,1,2,0,0}
-\frac{t}{8 \eps ^2}\BasisI{0,2,0,0,1,0,2,0,0}
%%%%% end : Idb4
\,,
\end{aligned} 
\label{CanonicalBasisExamples}
\end{equation}
but the unwanted singular coefficients are still present.

We would like to find a basis where reductions of integrals
have no singular coefficients.  We turn to that question in
the next section.

\section{Obtaining a singularity-free integral basis}
\label{SingularityFreeAlgorithms}

\def\basiccoeff{C}
\def\coeff#1{\basiccoeff^{[#1]}}
\def\trunccoeff#1{\widetilde \basiccoeff^{[#1]}}
\def\basicepscoeff{d}
\def\epscoeff#1{\basicepscoeff^{[#1]}}
\def\removedcoeff#1{U^{[#1]}}
\def\int#1{I^{[#1]}}
\def\basiccount{N}
\def\count#1{\basiccount^{[#1]}}
\def\basicibpcount{M}
\def\ibpcount#1{\basicibpcount^{[#1]}}
\def\rowred#1{R^{[#1]}}
\def\process#1{P^{[#1]}}
\def\orderlist#1{l^{[#1]}}
\def\mastercount{B}
\def\perm#1{S^{[#1]}}
In this section, we present two algorithms for
constructing a singularity-free integral basis. 
The first is based on the idea of
sequential $D=4$ IBP reduction; the second on 
Gaussian elimination in a local ring in $\eps$, that is
avoiding division by $\eps$. Given IBP relations for all 
Feynman integrals appearing in a given 
scattering process,  our algorithms provide a 
singularity-free basis for the process.

In the standard Laporta algorithm \cite{Laporta:2000dsw}, 
given $\basicibpcount$ IBP relations in 
$\basiccount$ Feynman integrals, 
one fixes the ordering of integrals, 
namely $I_1,\ldots I_{\basiccount}$, 
based on the integral sector, 
denominator degree, and numerator degree. Complicated 
integrals in the sense of Laporta will have smaller-valued 
subscripts.  This will order them at the beginning of
the list, and thereby target them for removal by 
row reduction.
We can organize the coefficients of the integrals in the
$\basicibpcount$ IBP relations
as an $\basicibpcount\times \basiccount$ 
matrix $\basiccoeff$, 
whose $j^{\textrm{th}}$ column 
corresponds to the integral $I_j$. 

The IBP system could in principle be generated through a 
Laporta-type
algorithm~\cite{Laporta:2000dsw}, using a template of 
symbolic IBPs~\cite{Lee:2013mka}, or by other means.  
In this article,
we use IBP relations obtained using
a generating-vector approach as implemented in
the \textsf{NeatIBP\/} \cite{Wu:2023upw,Wu:2025aeg} package.  The generating-vector
approach, as discussed earlier, eliminates integrals with
doubled propagators at the very start.  It also provides a 
minimal set of relations, for which 
$\basiccount - \basicibpcount = \mastercount$, where $\mastercount$ 
is the number of basis integrals.  
The approach in~\textsf{NeatIBP\/} is a natural one
when a basis free of doubled propagators
is desirable (which is almost always the case). 
These features are not
essential to the working of the algorithm, but do make using
it more efficient.  
The number of master integrals is 
of course independent of the algorithm used to 
generate the IBP equations.  

Should a different algorithm be used, the
generated set of IBP relations may be redundant; we
assume that redundant relations have been removed before
using the algorithms described below.  In addition,
the initial IBP equations may not suffice to
reduce `boundary' integrals (typically at or above the maximal
degree in loop momenta in the list of target integrals),
but will only suffice to eliminate them from the system.
We also assume that this elimination has been done, so
that we are left only with target integrals, those
whose reductions we need.  (We include
all possible basis integrals in this set.)
The initial IBP relations are necessarily linear in
$\eps$, but the removal of `boundary' integrals may raise
the degree in $\eps$.  The algorithms do not assume linearity
in $\eps$, though some steps will be simpler if the
IBP system remains linear.

Implementations of the algorithms described in this article
are available at
\begin{center}
\url{https://github.com/StefanoDeAngelis/SingularityFree}.
\end{center}

 \subsection{Reduction by Projection}

Our starting point is a system of IBP equations for Feynman
integrals,
\begin{equation}
\label{eq:neatIBPs}
   \sum_{j = 1}^{\count0} \coeff0_{ij} 
   \int0_j = 0\ ,\quad i = 1,\ldots,\ibpcount0\ ,
\end{equation}
where $\count0$ 
is the initial number of integrals appearing in the IBP relations.
We have added a superscript as the values in this algorithm
will change at every iteration.

The list of integrals $\int0_j$ is initially sorted 
according to Laporta's integral ordering.  During the algorithm,
we will re-sort the integrals.  It is possible to keep
track of the reduction coefficients to the basis we are
constructing, either analytically or using finite-field
reconstruction~\cite{Peraro:2019svx,Klappert:2020aqs,%
Magerya:2022hvj} 
with rational values for the Mandelstam invariants.  However,
it is more efficient to make use of a prior reduction to
a Laporta basis, and find the change of basis to the new
singularity-free basis afterwards.  Accordingly, we focus on
finding the integrals in the new basis.  
We therefore choose rational values for all 
Mandelstam invariants\footnote{We must take care that
these rational values be generic and in particular not yield a
\textit{special kinematic\/} configuration, that is a kinematic 
configuration for which any of the 
\textit{letters\/}~\cite{Goncharov:2010jf,Dixon:2011nj,%
Dixon:2012yy}
of the Feynman integrals vanish.  For six-point or higher
integrals, we must also ensure that the chosen values satisfy
the Gram determinant identities enforcing the 
four-dimensionality of external momenta.}.

\def\Ord{\mathcal{O}}
As our goal is to find a basis where no coefficient 
has inverse
powers of $\eps$, we will use Gaussian 
elimination avoiding
all divisions by $\eps$.  There is one exception: 
if a row
is proportional to $\eps$, we may still divide it out.
Our algorithm is iterative; the superscripts 
in $\coeff{w}$,
$\int{w}$, and $\count{w}$ record the iteration number.
The algorithm proceeds as follows,
\begin{enumerate}
\item Initialize $w=0$.
\item \label{LoopStart} Define $\trunccoeff{w}$ via,
      \begin{equation}
        \begin{aligned}
        \trunccoeff{w}_{ij} &=  \coeff{w}_{ij} 
                                \big|_{\eps = 0}\,.
        \end{aligned}
      \end{equation}
Both of $\trunccoeff{w}$ and $\coeff{w}$ are 
$\ibpcount{w}\times\count{w}$ 
matrices.
The rank of $\trunccoeff{w}$ may 
be smaller than that of $\coeff{w}$, because in setting 
$\eps = 0$,
we are discarding combinations of linear relations 
which have 
overall factors of $\eps$.  The relations expressed
by $\trunccoeff{w}$ are not
exact, but hold up to corrections of $\Ord(\eps)$.

\def\redint#1{\overline I{\kern 2pt\kern -2pt}^{[#1]}}
\item \label{FirstGaussianElimination}
Perform Gaussian elimination on $\trunccoeff{w}$. 
Denote the independent integrals by $\int{w+1}_i$; they
are candidate basis integrals.  This 
procedure yields a possibly \textit{overcomplete\/} 
set of master integrals, with 
$i = 1, \dots , \count{w+1} \le \count{w}$.  
It may be overcomplete because
we have discarded combinations of relations 
proportional to $\eps$  in step \ref{LoopStart}.
We 
will denote by $\redint{w}_j$ the integrals 
which are reducible using the result of this
Gaussian elimination. The expressions for these
reducible integrals obtained here will of course
only be valid up to corrections\footnote{This can 
include reductions 
setting integrals to zero, where the true expression 
for the integral may simply be $\eps\times \int{w+1}_r$, for 
example.} of $\Ord(\eps)$.  

\item[\ref{FirstGaussianElimination}${}^\prime$.] 
(optional, preparatory to step \ref{SecondGaussianElimination}${}^\prime$) 
If using the alternative
step~\ref{SecondGaussianElimination}${}^\prime$ and
if $\coeff{w}$ is linear in $\eps$, extract the terms linear
in $\eps$,
\begin{equation}
    \epscoeff{w}_{ij} =  \frac{1}{\eps}\bigl(\coeff{w}_{ij}-\trunccoeff{w}_{ij}\bigr) \,.
\end{equation}
If $\coeff{w}$ is not linear in $\eps$, construct and retain 
the row-reduction matrix $\rowred{w}$, which carries out the 
Gaussian elimination on $\trunccoeff{w}$ by left multiplication. For example, this can be achieved using Gauss-Jordan elimination.

\item \label{TerminationCondition}
Check if $\count{w+1} = \mastercount$. If so, we have 
obtained a new basis, and the algorithm terminates.
If $\count{w+1} > \mastercount$, continue with the next step.

\item We reorder the integrals so that the independent 
integrals obtained in step~\ref{LoopStart} are ordered last.
We do this by defining a permutation matrix $\perm{w}$
such that,
\begin{equation}
 (\perm{w})^{\vphantom[}_{i j}\int{w}_j =  
 (\redint{w}, \int{w+1})_i\ .
\end{equation}
The orderings within the $\redint{w}$ and $\int{w+1}$ 
subsets are left unchanged; this defines $\perm{w}$ 
uniquely.
The IBP relations must be permuted by the inverse
permutation: $\coeff{w} (\perm{w})^{-1})$, so
we are left with the following set of IBP 
relations to consider,
 \begin{equation}
\sum_{j,k = 1}^{\count{w}} \coeff{w}_{i k} 
((\perm{w})^{-1})_{k j} (\redint{w}, \int{w+1})_j = 0\ ,
 \quad i = 1,\ldots,\ibpcount{w}\,.
\end{equation}
This set is equivalent to the original set of
IBP relations~\eqref{eq:neatIBPs}, once the previously
removed relations 
$\removedcoeff{0},\ldots,\removedcoeff{w-1}$ (defined below) are
added back.

\def\reduced#1{r^{[#1]}}
\item \label{SecondGaussianElimination}
We now perform Gaussian elimination on 
$\coeff{w} (\perm{w})^{-1}$.  The resulting
matrix still has $\ibpcount{w}$ rows.  
(Were we to have started with
IBP relations without removing redundant equations, 
we would have zero
rows here which we would then have to discard.)  
We can identify four 
submatrices in it,
\newcommand{\topbraceA}{%
  \overbrace{%
\hphantom{\begin{matrix}\phantom{00}&\phantom{00}&\dots&
             \phantom{00}\end{matrix}}%
  }^{\text{\normalsize $\count{w}-\count{w+1}$}}%
}
\newcommand{\topbraceB}{%
  \overbrace{%
    \hphantom{\begin{matrix}0&0&0&0&0\end{matrix}}%
  }^{\text{\normalsize $\count{w+1}$}}%
}
\newcommand{\leftbraceA}{
  \text{$\count{w}-\count{w+1}$}\left\{\vphantom{\begin{matrix}0\\0\\\ddots\\0\end{matrix}}\right.
}
\newcommand{\leftbraceB}{
  \text{$\count{w+1}-\mastercount$}\left\{\vphantom{\begin{matrix}0\\0\\0\\0\end{matrix}}\right.
}
\begin{equation}
\settowidth{\dimen1}{$\leftbraceA$}
\settowidth{\dimen2}{$\topbraceB$}
\begin{matrix}
\begin{matrix}
\hspace{\dimen1}&\topbraceA&\topbraceB& 
\end{matrix}
\\[-0.5ex]
\begin{matrix}
\hfill\leftbraceA \\
\hfill\leftbraceB
\end{matrix}
\hspace{-0.35em}
\begin{bmatrix}
\begin{matrix}
1 & 0 & \dots & 0 \\
 & 1 & \dots & 0 \\
 &  & \ddots & \,0\, \\
\phantom{00} & \phantom{00} & \phantom{00} & 1
\end{matrix}
& \makebox[\dimen2]{\LARGE$*$} \\
\vphantom{\begin{matrix}0\\0\\0\\0\end{matrix}} 
\text{\large$0$} 
& \text{\large$\reduced{w+1}$}
\end{bmatrix}
&
\hspace{-0.5em}
\begin{bmatrix}
\vphantom{\begin{matrix}0\\0\\ \ddots \\0\end{matrix}} \text{\large$\redint{w}$} \\
\vphantom{\begin{matrix}0\\0\\0\\0\end{matrix}} \text{\large$\int{w+1}$}
\end{bmatrix}
\text{\large$= 0\ .$}
\end{matrix}
\end{equation}
The upper-left submatrix is the identity;
the upper-left and upper-right parts of the matrix 
together reduce the integrals in $\redint{w}$
to linear combinations of $\int{w+1}$s.  
These parts together form the matrix $\removedcoeff{w}$.
 The lower-left part of the matrix is
identically zero.  So long
as the set of $\int{w+1}$s is overcomplete, the lower-right
part of the matrix will be nontrivial and will provide
further identities between them.  In general, its
elements may be divergent as $\eps\rightarrow0$.  We denote
the most negative power of $\eps$ in a given row
$\count{w}-\count{w+1}<j\le \count{w}-M$ by $-h_j$.
We define the $(\count{w+1}-\mastercount)\times\count{w+1}$
coefficient matrix $\coeff{w+1}$ used in the next iteration
in terms of the lower-right hand matrix, removing the
singularities in $\eps$.  We can formally define it 
as follows,
\begin{equation}
\coeff{w+1}_{ij} = 
\eps^{h_i} 
\reduced{w+1}_{i+\delta,\,j+\delta}\,,
\qquad 1\le i\le \count{w+1}-\mastercount\,,
1\le j\le \count{w+1}\,.
\end{equation}
where $\delta=\count{w}-\count{w+1}$
This matrix is now nonsingular as $\eps\rightarrow0$.
We have now obtained a new, smaller, IBP system 
for the $\int{w+1}$,
\begin{equation}
   \sum_{j = 1}^{\count{w+1}} \coeff{w+1}_{ij} 
   \int{w+1}_j = 0\ ,\quad i = 1,\ldots,\ibpcount{w+1}\ ,
\end{equation}
This description assumes a symbolic treatment of $\eps$.
As we have chosen rational values for the Mandelstam
invariants, it is the only symbol left.  We could
speed up this step by choosing finite-field values for
the Mandelstam invariants, using a set of
different values for $\eps$, and using finite-field
reconstruction techniques to find the (rational) functional
dependence on $\eps$.

\def\leastcommoneps#1{P^{[#1]}}
\item[\ref{SecondGaussianElimination}${}^\prime$.] 
(alternative to step \ref{SecondGaussianElimination})  Gaussian elimination on $\coeff{w} (\perm{w})^{-1}$, with generic $\eps$-dependent entries, can be computationally expensive. We can instead perform this alternative step.
If $\coeff{w}$ is linear in $\eps$, we can form a new
$\basicibpcount\times 2\count{w}$ matrix by placing
$\epscoeff{w}$ to the right of $\trunccoeff{w}$ after
right-transforming both by $(\perm{w})^{-1}$,
\begin{equation}
    \bigl(\trunccoeff{w}(\perm{w})^{-1}\,\big|\,
    \epscoeff{w}(\perm{w})^{-1}\bigr)\,,
\end{equation}
and performing Gaussian elimination on it. We then rebuild
$\coeff{w+1}$ by examining each row, with $v_1$ given by the
first $\count{w}$ elements, and $v_2$ by the remaining
$\count{w}$ elements.  The output row is,
\begin{equation}
    \begin{aligned}
        &v_1 + \eps v_2\,,\qquad & v_1\neq 0\,,\\
        & v_2\,,\qquad & v_1 = 0\,.
    \end{aligned}
\end{equation}
If $\coeff{w}$ is not linear in $\eps$, we instead construct the matrix,
\begin{equation}
\label{eq:reduced_shuffled}
    \rowred{w} \coeff{w} (\perm{w})^{-1} \ .
\end{equation}
The last $\count{w+1} - \mastercount$ rows of this matrix 
will all be proportional to positive powers of $\eps$ by 
construction. Indeed, if we had used $\trunccoeff{w}$ in 
\eqn{eq:reduced_shuffled}, rather than $\coeff{w}$, such rows 
would be identically zero.  We can now build the diagonal
matrix $\leastcommoneps{w}$, whose entries are 
\begin{equation}
    \leastcommoneps{w}_{i i} = \frac{1}{\eps^{l_i}} \ , \quad \textrm{and} \quad \leastcommoneps{w}_{i j} = 0 \ \textrm{for} \ i \neq j\ ,
\end{equation}
where $l_i$ is the \textit{lowest overall power} of $\eps$ 
in $i^{\text{th}}$ row. Finally, we define
\begin{equation}
    \coeff{w+1} = \leastcommoneps{w} \rowred{w} \coeff{w} (\perm{w})^{-1}\ .
\end{equation}

\item We increment the iteration number, and continue with
step \ref{LoopStart} for the new coefficient matrix and set of
integrals. 
\end{enumerate}

The algorithm terminates at step 
\ref{TerminationCondition}, when
the new basis is no longer overcomplete. 
At that iteration, the set of master integrals 
is the desired singularity-free basis.

\subsection{Gaussian Elimination in a Local Ring}
\label{LocalRingAlgorithm}

Our second algorithm performs Gaussian elimination in
a \textsl{local ring\/}. Formally, a local ring $R\equiv \mathbb F[x]_{\mathbf m}$ is a polynomial ring $\mathbb F[x]$ {\it localized} on a maximal ideal $\mathbf m$,
\begin{equation}
  R=\bigg\{  \frac{f(x)}{g(x)}, \quad f(x), g(x) \in \mathbb F[x] , g(x)\not \in \mathbf m \bigg\}.
\end{equation}
In other words, the local ring $R$ consists of rational 
functions whose denominators are {\it not} in $\mathbf m$. We 
refer to the standard textbook \cite{CoxLittleOShea} for the 
details about the concept and related computations in local 
rings.

In our application, we use the local ring 
$\mathbb Q[\epsilon]_{\langle \epsilon\rangle}$. The maximal 
ideal $\langle\epsilon \rangle$ is the set of all polynomials 
in $\epsilon$ vanishing at $\epsilon=0$. The ring 
$\mathbb Q[\epsilon]_{\langle \epsilon\rangle}$ consists of 
all rational functions in $\epsilon$ which are finite at $\epsilon=0$.  We consider the Gaussian elimination of IBP identities, and demand that all the coefficients in the intermediate steps and the final results are in the local ring $\mathbb Q[\epsilon]_{\langle \epsilon\rangle}$. This forbids division
by $\eps$, while division by polynomials that 
are finite at $\eps=0$ is allowed.

This method has the advantage that this form of Gaussian
elimination is very similar to the 
standard one. Available Gaussian elimination packages can 
easily be adapted for our purpose with minor adjustments.

Using the standard Laporta ordering, the 
Gaussian elimination of $\coeff0$
(without any column swaps) would implement IBP
reduction and give us the set of master integrals.
However, in general, this process may yield poles
in $\eps$ in the reduction coefficients.  This happens
because pivots proportional to $\eps$ will result in
division by $\eps$ and thence poles in $\eps$ further
along in the pivot row.  A pivot proportional to 
a power of $\eps$
will avoid generating poles only if all other row elements
are also proportional to the same power of $\eps$.

In order to avoid generating poles in $\eps$, we must
avoid pivots proportional to $\eps$ (after removing
common factors of $\eps$).  Could we simply forbid terms 
proportional to $\eps$ in selecting a pivot? In general, 
when we disallow column swaps, all pivot candidates are
in the leading column past already-reduced rows.
If all nonzero entries in this column are proportional to 
$\eps$, then forbidding such pivots will cause 
the Gaussian elimination procedure to terminate too early.
The IBP reduction will then be incomplete.

Instead, we modify the standard procedure as follows: 
we allow column swaps and forbid pivots proportional to 
$\eps$. At each step of the Gaussian elimination, we divide
out any overall power of $\eps$ in every remaining row.
We then swap the columns to select a pivot which is not proportional to $\eps$.  We keep track of these swaps,
as they will lead to a new choice of master integrals.
In general, it will be different from that obtained by
the usual Laporta procedure.

As in the sequential algorithm described in the previous
subsection, we start with a set of integrals ordered using
the Laporta ordering, with `complicated' integrals ordered
earlier than `simple' ones.

The steps of the algorithm are as follows,
\begin{enumerate}
\item Define an index $p$ and initialize it, $p=1$;
define a list $\orderlist{1}$ containing the relevant
sorted integrals, 
$\orderlist{1}=[I_1,I_2,\ldots I_\basiccount ]$. 
As above, the matrix $\coeff{1}$ contains the coefficients
of the integrals in the
$N$ input IBP relations.  The index $p$ is also  the iteration counter.

\item \label{LocalRingLoopStart}
Divide each row $i$ ($p\leq i\leq \basicibpcount$) of $\coeff{p}$ by the greatest common
divisor (gcd) of the row's elements.

\item \label{LocalRingSearch}
Find a candidate pivot: scan all elements 
of $\coeff{p}$, columns $v$ from the $p^{\rm th}$ rightwards, rows $u$ from the $p^{\rm th}$
to bottom within each column.  

Select $(u,v)$ such that (a) $\coeff{[p]}_{uv}$ is 
nonzero and \textit{not\/} proportional to $\eps$ (b) $\orderlist{p}_v$ is the integral with the {\it smallest subscript} (that is,
the earliest position in the original list
$\orderlist{1}$) (c) the $u^\textrm{th}$ row is the sparsest 
row (or the first of a set of
equally sparsest rows). This corresponds to
a so-called `minimal swap' strategy.  Other strategies
are possible, and will be discussed below.

\item Pivot: swap the $p^\textrm{th}$ and $u^\textrm{th}$
rows and the $p^\textrm{th}$ and $v^\textrm{th}$ columns 
of $\coeff{[p]}$.
Swap the $p^\textrm{th}$ and $v^\textrm{th}$ element 
of $l^{[p]}$ to get a new list $l^{[p+1]}$.

\item \label{LocalRingNormalize}
Normalize: divide the $p^\textrm{th}$ row of
$\coeff{p}$ by $\coeff{p}_{pp}$.

\item Reduce: subtract appropriate multiples of the
$p^\textrm{th}$ row of $\coeff{p}$ to eliminate all leading
entries of rows $p+1,\ldots,\basicibpcount$. Denote the 
resulting new matrix by $\coeff{[p+1]}$. 

\item Increment $p$ by one. 
If $p\leq \basicibpcount$, continue at 
step~\ref{LocalRingLoopStart}.
\end{enumerate}

The resulting choice of master integrals is given
by the last $\basiccount-\basicibpcount$ elements in the 
list $I^{\basicibpcount+1}$. 
By construction, the reductions
of all other integrals have nonsingular coefficients.
We demonstrate the correctness of this algorithm in
Appendix~\ref{AlgorithmProof}.

At the search step (step~\ref{LocalRingSearch}), 
there are other 
possible criteria for choosing a pivot.  Indeed,
if we first assemble a list of all nonzero elements
not proportional to $\eps$ in the submatrix at or below
the $p^{\rm th}$ row and at or to the right of the
the $p^{\rm th}$ column, the list will generally contain more than one element.
The `minimal-swap' strategy described above 
attempts to find the indices of the most `complicated'
integral in the sense of the original ordering, so
long as the candidate pivot is not proportional to
$\eps$.  That is, the goal is to obtain a 
singularity-free basis which is as close as possible
to the Laporta basis.

In addition to this, we have also examined 
what we term a `Markowitz' strategy \cite{Markowitz}. This strategy
attempts to preserve the sparsity of the matrix
during Gaussian elimination.  
We define it as follows: denote by $r_i$ the number of
nonzero entries in row $i$, and by $f_j$ the number of 
nonzero entries in column $j$. 
Define the Markowitz numbers $m_{ij} = (r_i-1)(f_j-1)$.
For all nonzero elements $\coeff{p}_{ij}$ ($i\geq p$ 
and $j\geq p$) which are not proportional to $\eps$, 
choose the one with minimal Markowitz number to be the pivot.

In practice, we believe the `minimal-swap' strategy
is preferable.   As in the iterative-reduction algorithm
discussed in the previous section, when our sole purpose
is to obtain the set of basis integrals,
it is helpful to set all Mandelstam 
variables and masses to generic rational number values.

\section{Examples}
\label{ExamplesSection}

In this section, we apply the algorithms presented above to 
two integral families: the two-loop massless 
planar and nonplanar double boxes.  In both cases, we start
with all (`generic') integrals with up to the fourth 
power of ISPs alongside the set of finite integrals
determined in ref.~\cite{Gambuti:2023eqh}\footnote{We omit the
integrals that vanish upon symmetrization.}.  We
start with the IBP system for the generic integrals
as found using 
generating-vector method as implemented in \textsf{NeatIBP}.
We also construct a set of relations between the generic 
and finite integrals obtained by expanding the numerators of the
finite integrals.
As described in \sect{SingularityFreeAlgorithms}, 
we sort the generic integrals with decreasing  
Laporta weight.  When adding the set of finite integrals,
we place them at the very end, effectively assigning them
the lowest Laporta weights.
This fixes the matrix of integral 
coefficients $\coeff{0}$.

\subsection{Two-loop Planar Double Box}

We begin by running both algorithms solely on the 
generic integrals.  We find the same set of master
integrals,
\begin{equation}
\begin{aligned}
\{&
\BasisI{0,1,1,1,1,0,1,0,0},\BasisI{1,0,1,1,1,0,1,0,0},
\BasisI{1,1,0,1,1,0,1,0,0},\BasisI{1,1,1,1,0,1,1,0,0},\\
&\BasisI{1,1,1,1,1,0,1,0,0},\BasisI{1,1,1,1,1,1,0,0,0},\BasisI{1,1,1,1,1,1,1,-1,0},\BasisI{1,1,1,1,1,1,1,0,0}
\}\,.
\end{aligned}
\label{FirstSingularityFreeBasis}
\end{equation}
Let us look at the decomposition of the two
example integrals from 
\eqns{LaportaBasisExamples}{CanonicalBasisExamples}
into this new basis,
\begin{equation}
\begin{aligned}
\db[\Den_6^2] &= 
%%%%% begin : Idb5
\frac{\eps (2 s+t)}{s (1-2 \eps )} 
\BasisI{0,1,1,1,1,0,1,0,0} 
-\frac{\eps (3 s+t)}{s (1-2 \eps )} 
\BasisI{1,1,0,1,1,0,1,0,0}
\\&+\frac{(6 s+2 t \eps )}{6(1-2 \eps )} 
\BasisI{1,1,1,1,1,0,1,0,0}
-\frac{2 s \eps}{3 (1-2 \eps )} \BasisI{1,1,1,1,0,1,1,0,0}
\\&+\frac{2 s \eps}{3 (1-2 \eps )}
\BasisI{1,1,1,1,1,1,0,0,0}
+\frac{3 \eps}{1-2 \eps }\BasisI{1,0,1,1,1,0,1,0,0}
%%%%% end : Idb5
\,,
\\
\db\biggl[\Gram{\loopm_1&1&2}{\loopm_2&3&4}\biggr] &= 
%%%%% begin : Idb6
-\frac{s^2 (3+2 \eps)}{24 (1-2 \eps)}
\BasisI{1,1,1,1,0,1,1,0,0}
-\frac{s^2 (3-8 \eps)}{24 (1-2 \eps)}
\BasisI{1,1,1,1,1,1,0,0,0}
\\&+\frac{s^2}{8}\BasisI{1,1,1,1,1,1,1,-1,0}
+\frac{s+t}{2} \BasisI{1,1,0,1,1,0,1,0,0}
\\&-\frac{s (1-6 \eps)}{4 (1-2 \eps)}
\BasisI{1,0,1,1,1,0,1,0,0}
-\frac{t}{2} \BasisI{0,1,1,1,1,0,1,0,0}
%%%%% end : Idb6
\,.
\end{aligned}    
\label{eq:finite_integral_with_finite_coefficients}
\end{equation}
Unlike the reductions of these two integrals given
in \eqns{LaportaBasisExamples}{CanonicalBasisExamples},
the coefficients of the reductions here are not
singular as $\eps\rightarrow0$.  (The basis integrals 
appearing here still
are singular.)  These reductions are examples of why
the basis in \eqn{FirstSingularityFreeBasis} is
a singularity-free basis.

We can go further.  The second integral is one of the finite
integrals constructed in ref.~\cite{Gambuti:2023eqh};
we can
consider including it as a basis integral in place of
one of the integrals in \eqn{FirstSingularityFreeBasis}.
To implement this improvement in the basis, add
the set of finite integrals at the end of list of
generic integrals, and also add the relations expressing
them in terms of generic integrals to the list
of IBP identities.
The placement at the end effectively assigns them lower
Laporta weights, favoring their appearance over
divergent integrals.  We then
run the algorithms on this extended set of integrals.
Both yield the same new basis in which the 
two $\Ord(1/\eps^{4})$-divergent double-box integrals
(the two last entries in \eqn{FirstSingularityFreeBasis})
are replaced by the two finite double boxes 
with degree-two numerators,
\begin{equation}
\begin{aligned}
F_1 &= \db\biggl[\Gram{\loopm_1&1&2}{\loopm_2&3&4}\biggr]\,,\\
F_2 &= \db\biggl[\Gram{\loopm_1&1&2&3}
                      {\loopm_2&1&2&3}\biggr]\,.
\end{aligned}
\end{equation}
For the planar
double box, this replacement
retains the finiteness of the integral 
coefficients. The resulting basis is,
\begin{equation}
\begin{aligned}
\{& F_1\,, F_2\,,
\BasisI{1,1,1,1,0,1,1,0,0}\,,
\BasisI{1,1,1,1,1,0,1,0,0}\,,
\BasisI{1,1,1,1,1,1,0,0,0}\,,\BasisI{1,0,1,1,1,0,1,0,0}\,,\\
&\BasisI{1,1,0,1,1,0,1,0,0}\,,\BasisI{0,1,1,1,1,0,1,0,0}\,\}\,.
\end{aligned}
\end{equation}
These integrals turn out to be the same as 
the $s$-channel integrals
in the basis simplifying the presentation of
the two-loop four-gluon all-plus amplitude in 
ref.~\cite{Gambuti:2023eqh}.

\subsection{Two-loop Nonplanar Double Box}

\def\np{\textrm{NP}}
We can repeat the construction for the nonplanar
double box.  The expression for this integral
is given by \eqn{eq:feynman_integral} with 
$\Den_j$ replaced by $\Den_j^{\np}$ where,
\begin{equation}
\begin{aligned}
\Den_1^\np&=\loopm_1^2\,,\quad
\Den_2^\np=(\loopm_1-k_1)^2\,,\quad
\Den_3^\np=(\loopm_1-K_{12})^2\,,\quad
\Den_4^\np=\loopm_2^2\,,\\
\Den_5^\np&=(\loopm_2+k_4)^2\,,\quad
\Den_6^\np=(\loopm_2-\loopm_1+K_{124})^2\,,\quad
\Den_7^\np=(\loopm_1-\loopm_2)^2\,.
\end{aligned}
\end{equation}
The two ISPs here are,
\begin{equation}
    \ISP_1^\np = (\loopm_1+k_4)^2\,,\quad
    \ISP_2^\np = (\loopm_2+k_1)^2\,.
\end{equation}

The Laporta basis for this integral and its descendants is,
\begin{equation}
    \begin{aligned}
        \{
        &\BasisI{ 0, 0, 1, 1, 0, 0, 1, 0, 0}, 
        \BasisI{ 0, 1, 0, 0, 1, 0, 1, 0, 0}, 
        \BasisI{ 0, 1, 0, 1, 0, 1, 0, 0, 0}, 
        \BasisI{ 1, 0, 1, 0, 1, 0, 1, 0, 0},\\ 
        &\BasisI{ 0, 1, 0, 1, 1, 1, 1, 0, 0}, 
        \BasisI{ 0, 1, 1, 1, 0, 1, 1, 0, 0}, 
        \BasisI{ 0, 1, 1, 1, 1, 0, 1, 0, 0}, 
        \BasisI{ 1, 1, 1, 0, 1, 0, 1, 0, 0},\\ 
        &\BasisI{ 1, 1, 1, 1, 0, 1, 0, 0, 0}, 
        \BasisI{ 1, 0, 1, 1, 1, 1, 1, 0, 0}, 
        \BasisI{ 1, 1, 1, 1, 1, 1, 1, 0, 0}, 
        \BasisI{ 1, 1, 1, 1, 1, 1, 1, -1, 0}
        \}\,.
    \end{aligned}
\end{equation}

We run both algorithms on the list
of generic integrals supplemented by the finite
integrals found in ref.~\cite{Gambuti:2023eqh},  Both again
yield the same basis,
\begin{equation}
\begin{aligned}
\{
&\BasisI{0, 1, 0, 1, 1, 1, 1, 0, 0}, 
\BasisI{0, 1, 1, 1, 1, 0, 1, 0, 0}, 
\BasisI{1, 0, 1, 1, 1, 1, 0, 0, 0},
 \BasisI{1, 1, 0, 1, 1, 1, 0, 0, 0}, \\
& \BasisI{0, 1, 1, 1, 1, 1, 0, 0, 0}, 
\BasisI{1, 0, 0, 1, 1, 1, 1, 0, 0},
\BasisI{1, 0, 1, 1, 1, 1, 1, 0, 0}, 
 \BasisI{1, 1, 0, 1, 1, 0, 1, 0, 0}, \\
&\BasisI{1, 1, 1, 1, 1, 0, 1, 0, 0}, 
\BasisI{1, 1, 1, 1, 1, 1, 0, 0, 0}, 
 \BasisI{1, 1, 1, 1, 1, 1, 1, 0, -1}, 
 \BasisI{1, 1, 1, 1, 1, 1, 1, 0, 0}\}\,.
\end{aligned}
\label{NonplanarDoubleBoxSingularityFreeBasis}
\end{equation}
In this case, no finite integrals survive in the basis.
The first four integrals in this basis diverge
as $1/\eps^2$ as $\eps\rightarrow 0$,
\begin{equation}
    \begin{split}
        \BasisI{0,1,0,1,1,1,1,0,0}&=\frac{\left[\log \left(1+\frac{1}{\chi }\right)-2 i \pi \right] \log \left(\frac{\chi }{1 + \chi}\right)}{2 s \epsilon ^2}+\Ord (\epsilon^{-1})\ ,\\
        \BasisI{0,1,1,1,1,0,1,0,0}&=\frac{\log ^2\chi +\pi ^2}{2 s(1 + \chi) \epsilon ^2}+\Ord (\epsilon^{-1})\ ,\\
        \BasisI{1,0,1,1,1,1,0,0,0}&=\frac{\pi ^2}{12 s \epsilon^2}+\Ord (\epsilon^{-1})\ ,\\
        \BasisI{1,1,0,1,1,1,0,0,0}&=\frac{[2 i \pi -\log (1 + \chi)] \log (1 + \chi)}{2 s \chi  \epsilon ^2}+\Ord (\epsilon^{-1})\,, 
    \end{split}
\end{equation}
where $\chi=t/s$ and the integrals have been evaluated 
in the $u$-channel, \textit{i.e.} at $s<0$ and $\chi > 0$.
The remaining eight integrals diverge as $1/\eps^4$
in the four-dimensional limit.

As an example, consider the reduction of the
following integral,
\begin{equation}
\begin{aligned}
&\hspace*{-10mm}\BasisI{1,1,1,1,1,1,1,-2,0} =\\
%%%%% begin : Indb1
&\frac{(t-2 s+4 t \eps -5 s \eps)}
{(1+4 \eps) (s+t)}
\BasisI{0,1,0,1,1,1,1,0,0}
+\frac{(s+2 t+7 s \eps +11 t \eps)}
{s (1+4 \eps)}
\BasisI{0,1,1,1,1,0,1,0,0} 
\\&
-\BasisI{1,0,1,1,1,1,0,0,0}
+\frac{t\left(s-2 t+4 s \eps -5 t \eps \right)}
{s (1+4 \eps) (s+t)}
\BasisI{1,1,0,1,1,1,0,0,0} 
\\&
+\frac{(3 s +2 t+9 s \eps+5 t \eps)}
{s (1+4 \eps)}
\BasisI{0,1,1,1,1,1,0,0,0} 
-\frac{3 (t+s \eps +4 t \eps)}{(1+4 \eps) (s+t)}
\BasisI{1,0,0,1,1,1,1,0,0} 
\\&
+\frac{(t-s)}{2}  \BasisI{1,0,1,1,1,1,1,0,0} 
-\frac{t\left(5 s+2 t +20 s \eps+11 t \eps\right)}
  {(1+4 \eps) s (s+t)}
\BasisI{1,1,0,1,1,0,1,0,0}
\\&
+\frac{(9 s-5 t)}{3} \BasisI{1,1,1,1,1,0,1,0,0}
+\frac{(4 t-11 s)}{6} \BasisI{1,1,1,1,1,1,0,0,0} 
\\&
+\frac{(2 t-3 s)}{2} \BasisI{1,1,1,1,1,1,1,0,-1} 
+\frac{t(t-s)}{2} \BasisI{1,1,1,1,1,1,1,0,0} 
%%%%% end : Indb1
\,.
\end{aligned}
\end{equation}
The coefficients are nonsingular as desired.  In contrast,
in the Laporta basis, some of the reduction coefficients
would be singular as $\eps\rightarrow0$.

The simplest application of this integral family is to the
subleading-color contributions to the two-loop all-plus 
four-gluon amplitude in pure Yang--Mills theory. 
The result for this amplitude was given in the 
millenium paper~\cite{Bern:2000dn},
\def\cb{I_{\textrm{NP}}}
\begin{equation}
\label{eq:allplus_nonplanar}
    \begin{split}
        A_{12 ; 34}^{\mathrm{NP}} = 
        i \frac{\spb1.2\spb3.4}{\spa1.2\spa3.4} s\, \Big(& \left(D_s-2\right) \cb\left[\lambda_{l_1-l_2}^2 \lambda_{l_2}^2+\lambda_{l_1-l_2}^2 \lambda_{l_1}^2+\lambda_{l_1}^2 \lambda_{l_2}^2\right]\\
        & +16\,\cb\left[\left(\lambda_{l_1-l_2} \cdot \lambda_{l_2}\right)^2-\lambda_{l_1-l_2}^2 \lambda_{l_2}^2 \right]\Big)\ ,
    \end{split}
\end{equation}
where $\cb$ is the nonplanar double box,
and $\lambda^\mu_{l}$ is the projection of the vector 
$\loopm^\mu$ on the subspace orthogonal to the 
external $D=4$ space spanned by the vectors $k_1^\mu$, 
$k_2^\mu$, $k_3^\mu$ and 
$v^\mu = {\epsilon^{\mu \nu \rho \sigma} k_{1 \nu}%
k_{2 \rho} k_{3 \sigma}}/{\sqrt{-G(1\, 2\, 3)}}$. 
In particular, we have
\begin{equation}
    \lambda_{p} \cdot \lambda_{q} = \eta_{\perp}^{\mu \nu} 
    p_{\mu} q_{\nu} \ ,
\end{equation}
where
\begin{equation}
    \eta_{\perp}^{\mu \nu} = \eta^{\mu \nu} 
    - \sum_{i=1}^{3} k_i^\mu \hat{k}_i^\nu + v^\mu v^\nu\,,
    \qquad \text{with} \qquad \hat{k}_i \cdot k_j = 
    \delta_{i j}\ .
\end{equation}
We can systematically rewrite the all-plus amplitude as a 
sum of basis integrals, each multiplied by a coefficient
rational in the invariants and $\eps$.
If we choose the singularity-free 
basis~\eqref{NonplanarDoubleBoxSingularityFreeBasis}, 
we can pull out an overall factor of $\eps$ in the 
expression for the amplitude.  The amplitude is then
manifestly no more divergent than $1/\eps^3$,
\begin{equation}
A_{12 ; 34}^{\mathrm{NP}} = 
i \frac{\spb1.2\spb3.4}{\spa1.2\spa3.4}\, \eps\,
\hspace*{-2mm}\sum_{I\in\textrm{basis}} c_{I} \, I\ .
\end{equation}
This contrasts with the apparent $1/\eps^4$ divergence
in the Laporta basis.  The would-be $1/\eps^3$ divergence
of course cancels between integrals, leaving an
overall $1/\eps^2$ divergence.

The appearance of a power of $\eps$ does give us
extra freedom to modify the basis. 
We can relax our requirement that all IBP reduction 
coefficients be free of poles in $\eps$ to the requirement
that they have no poles worse than $1/\eps$, and still end
up with amplitude expressions that are free of explicit
poles in $\eps$.  As it turns out, we can use this 
freedom to make the cancellations of the would-be $1/\eps^3$
divergences manifest.  The best candidates for new
elements of the basis are the finite nonplanar double box
integrals enumerated in ref.~\cite{Gambuti:2023eqh}.
We can trade five integrals with $1/\eps^4$ divergences
for finite integrals, four with degree-three numerators
and one with a degree-four numerator,
\begin{equation}
    \begin{aligned} 
        & \BasisI{1, 1, 1, 1, 1, 1, 1, -1, 0} \\
        & \BasisI{1, 1, 1, 1, 1, 1, 1, 0, 0} \\
        & \BasisI{0, 1, 1, 1, 1, 1, 0, 0, 0} \\
        & \BasisI{1, 1, 0, 1, 1, 0, 1, 0, 0} \\
        & \BasisI{1, 0, 0, 1, 1, 1, 1, 0, 0}
    \end{aligned}
    \quad
    \longrightarrow 
    \quad
    \begin{aligned} 
        & F_1^\np=\cb\biggl[\Gram{\loopm_1&1&2}{1&2&3} 
        \Gram{\loopm_2&4}{1&2} 
        \Gram{\loopm_1\!-\! \loopm_2& 3}{1& 2}\biggr]\,,\\
        & F_2^\np = \cb\biggl[
        \Gram{\loopm_2&4}{\loopm_1 \!-\! \loopm_2&3} 
        \Gram{\loopm_1& 1& 2}{1& 2& 3}\biggr]\,,\\
        & F_3^\np=\cb\biggl[\Gram{\loopm_1& 1&2}{1& 2& 3}
        \Gram{\loopm_2& 4}{2& 3} 
        \Gram{\loopm_1 \!-\! \loopm_2& 3}{2& 3}\biggr]\,,\\
        & F_4^\np=\cb\biggl[\Gram{\loopm_2& 4}{1& 2} 
        \Gram{\loopm_1& 1& 2}
             {\loopm_1 \!-\! \loopm_2& 3& 1}\biggr]\,,\\
        & F_5^\np=\cb\biggl[G(\loopm_1\,1\,2) 
        \Gram{\loopm_2& 4}{\loopm_1\!-\!\loopm_2& 3}\biggr]\,.
    \end{aligned}
\end{equation}
The choice of integrals on the left-hand side is based
on an examination of the pattern of cancellations of
the leading singularity in the expression for the
subleading-color terms in the amplitude.
One can check that the replacement of integrals on the 
left-hand side by those on the right results in 
integral-reduction coefficients whose worst divergence 
is as $1/\eps$. 
The resulting form of the subleading-color contributions is,
\begin{equation}
\label{eq:allplus_nonplanar_withfinite}
    \begin{split}
        \hspace*{-3mm}A_{12 ; 34}^{\mathrm{NP}} = 
        i \frac{\spb1.2\spb3.4}{\spa1.2\spa3.4}\, 
        \big\{
        & \eps\, c_1 \BasisI{0, 1, 0, 1, 1, 1, 1, 0, 0} 
         +\eps^2\, c_2 \BasisI{0, 1, 1, 1, 1, 0, 1, 0, 0}
        +\eps^2\, c_3 \BasisI{1, 0, 1, 1, 1, 1, 0, 0, 0}\\
        +&\eps^2\, c_4 \BasisI{1, 0, 1, 1, 1, 1, 1, 0, 0}
        +\eps^2\, c_5 \BasisI{1, 1, 0, 1, 1, 0, 1, 0, 0}
        +\eps^2\, c_6 \BasisI{1, 1, 0, 1, 1, 1, 0, 0, 0}\\
        +&\eps^2\, c_7 \BasisI{1, 1, 1, 1, 1, 1, 0, 0, 0}\\
        +&c_8 F_1^\np
        +c_9 F_2^\np
        +c_{10} F_3^\np
        +c_{11} F_4^\np
        +c_{12} F_5^\np\big\}\ .
    \end{split}
\end{equation}
We give explicit expressions for the coefficients $c_i$ in
Appendix~\ref{CoefficientsAppendix}.

\section{Conclusions}
\label{ConclusionsSection}

In this article, we have studied a new type of
basis for Feynman integrals, in which singularities in
$\eps$ are absent in the coefficients that emerge from
integral reductions.  All divergences then arise solely
from the integrals themselves.  
In \sect{SingularityFreeAlgorithms}, we presented two
independent algorithms for constructing such bases.  
We gave examples of their use in the systems
of planar and nonplanar double boxes and daughter integrals
in \sect{ExamplesSection}.  In the case of the planar
double box, a singularity-free basis is compatible with
choosing some basis integrals to be finite integrals,
as given for example in ref.~\cite{Gambuti:2023eqh}.  The basis
we find is similar to the modified basis already found
to simplify the expression of the leading-color
terms in the two-loop all-plus Yang--Mills
amplitude~\cite{Gambuti:2023eqh}.  
In the case of the nonplanar
double box, we cannot choose any basis integral to be
a finite integral while maintaining a fully singularity-free
basis.  Nonetheless, the use of a singularity-free basis
does simplify the expression for the subleading-color
contributions to the two-loop all-plus amplitude.  Trading
some of the basis integrals for finite integrals simplifies
the expression for the amplitude further.  It will be
interesting to extend these investigations to the
other helicity amplitudes. The results
described here
suggest that singularity-free bases can play a role in
simplifying expressions for scattering amplitudes, and that
it would be desirable to find a general criterion 
to balance judiciously
between the singularity-free criterion and the inclusion of
finite Feynman integrals in a basis of Feynman integrals.

\begin{acknowledgments}

We thank CERN, the GGI, MIAPbP and the KITP for 
their hospitality during different phases of the research
described here.
We thank Janko B\"ohm, Giulio Gambuti, Pavel Novichkov,
and Lorenzo Tancredi for helpful discussions.  DAK
also thanks Giulio Gambuti and Lorenzo Tancredi for an initial
collaboration on related topics.
This research was supported by the European Research
Council (ERC) under the European Union's research and
innovation program grant agreement 
ERC--AdG--885414 (`Ampl2Einstein');
in part by grant NSF PHY--2309135 to the Kavli 
Institute for Theoretical Physics (KITP). 
RM is funded by the Outstanding PhD Students Overseas 
Study Support Program of the University of Science 
and Technology of China. ZW is supported by 
the Hangzhou Human Resources and Social Security Bureau
through the First Batch of Hangzhou Postdoctoral Research 
Funding in 2024. YZ is supported by the NSF of China 
through Grant No.~12247103.

\end{acknowledgments}

\appendix

\section{Correctness of Local-Ring Algorithm}
\label{AlgorithmProof}

\noindent{\bf Claim} The Gaussian elimination algorithm given in  \sect{LocalRingAlgorithm}, for reduction over a local ring, generates the maximal number of pivots, the same as the rank of the original matrix of IBP identities. The output matrix $\basiccoeff$ is in row echelon form 
free of poles in $\eps$.

\noindent{\bf Proof}  In each iteration of the algorithm, 
before  step~\ref{LocalRingLoopStart}, for 
$p\leq i \leq \basicibpcount$, the $i$-th row must be 
nonzero, because the matrix has maximal row rank (recall that 
we require the IBP relations corresponding to the original 
matrix to be linearly independent); and a row elementary 
transformation does not change the rank of the matrix. Then
after step~\ref{LocalRingLoopStart}, the $i$-th row 
$p\leq i \leq \basicibpcount$ must contain at least one 
nonzero element which is not proportional to $\eps$. 
Therefore step~\ref{LocalRingSearch} is guaranteed to 
find a pivot not proportional to $\eps$. Accordingly the algorithm finds a pivot in each iteration. The total number of pivots would be $\basicibpcount$, which is also the rank of the matrix. Therefore the algorithm generates the maximal number of pivots. 

In each iteration, after pivoting, the algorithm normalizes the diagonal element to one. All the lower triangular terms are eliminated. Because the algorithm produces the maximal number of pivots, the resulting matrix is in row echelon form.

Because the pivot is not proportional to a power of $\eps$,
the normalization in step~\ref{LocalRingNormalize} does not 
create a pole in $\eps$. The row elementary transformations 
do not generate new poles in $\eps$. Because the input matrix 
has no pole in $\eps$, the output matrix is then free of 
$\eps$ poles. $\qed$

\section{Integral Coefficients for the Subleading-Color Contributions to the Two-Loop All-Plus Four-Point Amplitude}
\label{CoefficientsAppendix}

Here we list the expansions of the rational coefficients 
in \eqn{eq:allplus_nonplanar_withfinite} up to the 
required order in the $\eps$ expansion,
\begin{equation}
\begin{split}
 c_1 = & (D_s-2)\left[\frac{2 t  (s+t)}{3 s} -\frac{2  \left(3 s^4-11 s^3 t-17 s^2 t^2-12 s t^3-6 t^4\right)}{9 s^3} \epsilon \right]+\Ord(\epsilon^2 )\ ,\\
 c_2 = & (D_s-2) \left[-\frac{ 4 s^5+16 s^4 t+27 s^3 t^2+18 s^2 t^3+7 s t^4+2 t^5}{3 s^2 t (s+t)} \right]+\Ord(\epsilon )\ ,\\
 c_3 = & (D_s-2) \left[\frac{2 \left(11 s^2-2 t^2\right)}{3 s}\right]+\Ord(\epsilon )\ ,\\
 c_4 = & (D_s-2) \bigg[-\frac{1}{3} s^2 +\frac{\left(6 s^4-10 s^3 t-s^2 t^2+18 s t^3+9 t^4\right)}{18 t (s+t)} \epsilon\\
& + \frac{255
   s^4+203 s^3 t+680 s^2 t^2+954 s t^3+477 t^4}{108 t (s+t)} \epsilon ^2 \bigg] -12 s^2 \epsilon ^2+\Ord(\epsilon^3 )\ ,\\
 c_5 = & (D_s-2) \bigg[ \frac{4 t^2}{3 s} +\frac{2 t \left(6 s^4+22 s^3 t+31 s^2 t^2+18 s t^3+9 t^4\right)}{9 s^3 (s+t)} \epsilon\\
 & + \frac{ t \left(255
   s^4+619 s^3 t+1096 s^2 t^2+954 s t^3+477 t^4\right)}{27 s^3 (s+t)} \epsilon ^2\bigg] -\frac{48 t^2}{s} \epsilon ^2 +\Ord(\epsilon^3 )\ ,\\
 c_6 = & (D_s-2) \frac{ 6 s^5+22 s^4 t+29 s^3 t^2+10 s^2 t^3+3 s t^4+2 t^5}{3 s^2 t (s+t)}+\Ord(\epsilon )\ ,\\
 c_7 = & (D_s-2) \bigg[\frac{4}{9} (s+t)^2 +\frac{2 (s+t) \left(6 s^4+22 s^3 t+31 s^2 t^2+18 s t^3+9 t^4\right)}{27 s^2 t} \epsilon\\
 & + \frac{(s+t)
   \left(255 s^4+619 s^3 t+1096 s^2 t^2+954 s t^3+477 t^4\right)}{81 s^2 t} \epsilon ^2 \bigg]
   \\
   &-16 (s+t)^2 \epsilon ^2 +\Ord(\epsilon^3 )\ ,\\
 c_8 = & -(D_s-2) \frac{128 \left(2 s^4-3 s^2 t^2+2 s t^3+t^4\right)}{3 s^4 t^2 (s+t)^2}+\Ord(\epsilon )\ ,\\
 c_9 = & (D_s-2) \frac{512 \left(s^2+3 s t+3 t^2\right)}{3 s t^3 (s+t)^2}+\Ord(\epsilon )\ ,\\
 c_{10} = & (D_s-2) \frac{16 (2 s-t) (3 s+t) (s+2 t)}{3 s^3 t (s+t)}+\Ord(\epsilon )\ ,\\
 c_{11} = & -(D_s-2) \frac{64 \left(4 s^2-s t-t^2\right)}{3 s^3 (s+t)}+\Ord(\epsilon )\ ,\\
 c_{12} = & (D_s-2) \frac{32 t (s+t)}{3 s^4}
 +\Ord(\epsilon )\,.\\
\end{split}
\end{equation}
The overall factor $(D_s-2)$ echos the origin of 
each term in \eqn{eq:allplus_nonplanar}. The coefficients $c_{4,5,7}$ have $\Ord(\epsilon^4 )$ contributions 
not proportional to $(D_s-2)$, which cancel in the amplitude through $\Ord(\eps^0 )$.

\bibliography{finite}

\end{document}